\documentclass[aps,twocolumn,superscriptaddress,floatfix,prx,longbibliography]{revtex4-1}
\usepackage[utf8]{inputenc}
\usepackage{lipsum}
\usepackage{graphicx}
\usepackage{amsmath,amssymb}
\usepackage{braket}
\usepackage{mathtools}
\usepackage{hyperref}
\usepackage{multirow}
\usepackage{color}
\usepackage[normalem]{ulem}
\usepackage{amsfonts}
\usepackage{bbm}
\usepackage[ruled,vlined]{algorithm2e}
\usepackage{braket}

\renewcommand{\[}{\left[}
\renewcommand{\]}{\right]}

\makeatletter

\usepackage{xcolor}

\usepackage[normalem]{ulem} 

\usepackage[caption=false]{subfig}
\begin{document}


\title{Effective Luttinger parameter and Kane-Fisher effect in quasiperiodic systems}

\author{T.J. Vongkovit}
\affiliation{Department of Electrical and Computer Engineering, Princeton University, Princeton, NJ 08544, USA}
\author{Hansveer Singh}
\affiliation{Department of Physics, University of Massachusetts, Amherst, Massachusetts 01033, USA}
\author{Romain Vasseur}
\affiliation{Department of Physics, University of Massachusetts, Amherst, Massachusetts 01033, USA}
\author{Sarang Gopalakrishnan}
\affiliation{Department of Electrical and Computer Engineering, Princeton University, Princeton, NJ 08544, USA}


\begin{abstract}
The ground states of interacting one-dimensional metals are generically Luttinger liquids. Luttinger liquid theory is usually considered for translation invariant systems. The Luttinger liquid description remains valid for weak quasiperiodic modulations; however, as the quasiperiodic modulation gets increasingly strong, it is increasingly renormalized and eventually fails, as the system becomes localized. We explore how quasiperiodic modulation renormalizes the Luttinger parameter characterizing this emergent Luttinger liquid, using the renormalization of transmission coefficients across a barrier as a proxy that remains valid for general quasiperiodic modulation. We find, unexpectedly, that quasiperiodic modulation weakens the effects of short-range interactions, but enhances those of long-range interactions. We support the former finding with matrix-product numerics. We also discuss how interactions affect the localization phase boundary.

\end{abstract}

\maketitle


The nature of ground-state correlations in interacting disordered systems has been a question of longstanding interest~\cite{RevModPhys.70.1039, sanchez2010disordered, vojta2019disorder, RevModPhys.91.011002}. In one dimension, this question has been addressed from two complementary perspectives: perturbatively adding disorder to clean interacting systems~\cite{PhysRevB.37.325}, and perturbing the disordered system with weak interactions using the strong disorder renormalization group (SDRG)~\cite{PhysRevLett.100.170402}. These approaches agree in finding two phases---a localized insulating phase and a Luttinger-liquid phase---but give contrasting predictions for the critical behavior between them (see also Refs.~\cite{PhysRevLett.103.140402, PhysRevB.89.054204, PhysRevB.96.180202}). In many present-day experiments, the spatial modulations are quasiperiodic rather than random: both in cold-atom experiments using incommensurate potentials~\cite{roati2008anderson, ColdSchreiber2015, Sbroscia2020, 2016arXiv161207173L, lueschen_qp,  PhysRevLett.126.040603} and more recently in Moir\'e materials~\cite{mak2022semiconductor, fu2020magic}. 
Quasiperiodic systems exhibit many of the same phenomena as random ones---e.g., Anderson localization---but their properties are fundamentally different because of the deterministic, hyperuniform structure of quasiperiodic patterns~\cite{Chandran:2017ab, Crowley2018, Crowley2018a, PhysRevB.100.134206}. 
Quasiperiodic systems present distinctive theoretical challenges, as many of the methods used in the random case (such as the replica trick) are no longer useful. 

In the absence of controlled analytical methods, progress on understanding the ground-state properties of interacting quasiperiodic systems has come from a combination of numerical~\cite{vidal1999correlated, roux2008quasiperiodic, mastropietro2015localization, naldesi2016detecting, cookmeyer2020critical, Agrawal2019, PhysRevB.65.115114, PhysRevB.89.161106, PhysRevLett.126.036803, goncalves, PhysRevB.102.195142} and perturbative approaches. 
As in the random case, the two complementary perturbative approaches are to treat interactions nonperturbatively in the Luttinger-liquid framework, but perturb in disorder~\cite{vidal1999correlated}, or to begin with the scaling theory of the noninteracting critical point, and perturb in interactions, as in very recent work~\cite{goncalves}. These approaches yield distinct universality classes for the localization transition: in the former case, the transition is Lorentz-invariant; in the latter, the Luttinger liquid never emerges and the localization transition remains in the noninteracting universality class. 
To reconcile these pictures it is imperative to understand how Luttinger-liquid behavior emerges (or fails to emerge) for strong quasiperiodic modulation, and how Luttinger-liquid properties evolve with quasiperiodicity. We address these questions, focusing on interacting spinless fermions with finite-range interactions. (The spinful case was recently explored~\cite{murthy}.) 

Increasing the quasiperiodic potential strength causes the single-particle bands to get increasingly flat~\cite{PhysRevB.40.8225}. As flat bands are associated with enhanced interactions, so one might expect a quasiperiodic potential to strengthen interaction effects. Surprisingly, this is not the case in general; instead, the \emph{range} of the interactions plays a crucial part. For short-range interactions, the quasiperiodic potential suppresses interactions and make the system more free-fermion-like~\cite{goncalves}. On the other hand, long-range interactions are enhanced by the quasiperiodic potential. We establish these results by two distinct means. First, we perform density matrix renormalization group (DMRG) simulations of one-dimensional quasiperiodic spin chains~\cite{PhysRevLett.69.2863,SCHOLLWOCK201196} and extract the Luttinger parameter, which quantifies interaction strength. Second, we study of the renormalization of transmission through a barrier, adapting the treatment of Yue et al.~\cite{PhysRevB.49.1966} to the quasiperiodic case. The renormalization of the transmission coefficient gives us a way of extracting an effective Luttinger parameter from Hartree-Fock studies. 



\emph{Model}.---We will focus on a single-band model of spinless fermions, interacting with finite-range interactions, governed by the Hamiltonian 
\begin{equation}\label{eq1}
H = \sum_i (c^\dagger_i c_{i+1} + \mathrm{h.c.} + \lambda \cos(\varphi (i + \theta)) n_i) + \sum_{ij} V_{ij} n_i n_{j}.
\end{equation}
where $\varphi$ is the Golden ratio, $c_i$ annihilates a fermion on site $i$ and $n_i \equiv c^\dagger_i c_i$, and the overall energy scale has been set to unity.
We briefly review the properties of this model in the noninteracting case $V_{ij} = 0$. This is the familiar Aubry-Andr\'e (AA) model. When $\lambda < 2$ the single-particle eigenstates of this model are delocalized and ballistic; for $\lambda > 2$ all single-particle states are localized. Unusually, the localization transition occurs at the same value of $\lambda$ for all states in the spectrum. At the critical value $\lambda = 2$, not only are all the eigenstates localized, but the \emph{spectrum} is highly unusual: all the single-particle eigenvalues are in a fractal, measure-zero set, so the bands are anomalously flat~\cite{wilkinson, kohmoto1987critical}. For $\lambda < 2$ as one approaches the critical point, the spectrum consists of many increasingly flat minibands. The conventional wisdom is that these flat minibands will tend to make the system highly susceptible to interactions. 

We consider two choices for the interactions $V_{ij}$. First, for our exact numerical studies, we take the interactions to be nearest-neighbor, i.e., $V_{ij} = \Delta \delta_{i, j\pm 1}$. The nearest-neighbor model is appealing for numerical studies because it can be mapped via a Jordan-Wigner transformation to a canonical nearest-neighbor spin model, the anisotropic Heisenberg model. 
This allows for efficient DMRG simulations, and also allows us to compare our numerical results in the clean limit with exact results in the literature. However, as we will see, the range of the interaction plays an important part in the physics of the model. To allow us to tune this range $d$, we will also consider interactions that decay with a Gaussian envelope:
\begin{equation}\label{intrange}
    V_{\rm int}(|i - j|) = \frac{v_{\rm int}}{d\sqrt{\pi}}\exp{[-(i - j)^2/d^2]}.
\end{equation}
Long-range interactions can be naturally incorporated in our Hartree-Fock analysis. They are possible~\cite{zaletel} but potentially challenging to treat in DMRG. We note that the Hartree-Fock method has other advantages, as it allows us to address aspects of the dynamics (and disentangle distinct physical effects) in a way that would not be possible with DMRG.


When $\lambda < 2$, at a typical filling the Fermi level lies inside a miniband. One can treat sufficiently weak interactions by linearizing the single-particle spectrum about the Fermi level and treating the interactions projected onto a single miniband. This yields a Luttinger liquid, in which the strength of correlations is set by the Luttinger parameter $K$ (defined below). Linearizing becomes an increasingly inaccurate approximation as one approaches $\lambda = 2$: any finite interaction strength mixes many minibands. Nevertheless, we find that the effect of the other minibands manifests itself as a finite renormalization of the Luttinger parameter, without changing the asymptotic nature of correlations. 

One can understand our main results intuitively by thinking about the behavior of interactions projected onto a single miniband. The well-defined minibands at some $\lambda < 2$ are essentially the Bloch bands of a periodic approximant with denominator $q$ to the true incommensurate potential. As $q$ grows, the minibands split, so the Bloch states of each miniband are made up of increasingly well separated ($\sim q$) Wannier states. For any finite interaction range $d$, eventually $q \gg d$, so the interactions projected onto the miniband are contact-like. A contact interaction has no effect on identical spinless fermions, so the projected problem becomes noninteracting. This accounts for the surprising robustness of the flat minibands of the Aubry-Andr\'e model to interactions. 


\emph{DMRG results}.---We begin by extracting the Luttinger parameter directly from DMRG calculations on the ground state of the quasiperiodically modulated problem at half-filling. Our primary diagnostic for the Luttinger parameter is the variance of the charge in a half-system, i.e., the behavior of the following connected correlation function for a subsystem of length $\ell$:
\begin{equation}
F(\ell) = \left\langle \left(\sum_{i = 1}^{\ell}n_i\right)^2 \right\rangle - \left\langle \left(\sum_{i = 1}^{\ell} n_i \right)\right\rangle^2,
\end{equation}
where expectation values are taken in the ground state. In a Luttinger liquid with open boundary conditions in a system of size $L$, one expects $F(\ell) = \frac{K}{2\pi^2} \log( \frac{L}{\pi} \sin(\frac{\pi \ell}{L}))$, where $K$ is the Luttinger parameter~\cite{giamarchi_book}. 

\begin{figure}[tb]
    \centering
    \includegraphics[width=0.43\textwidth]{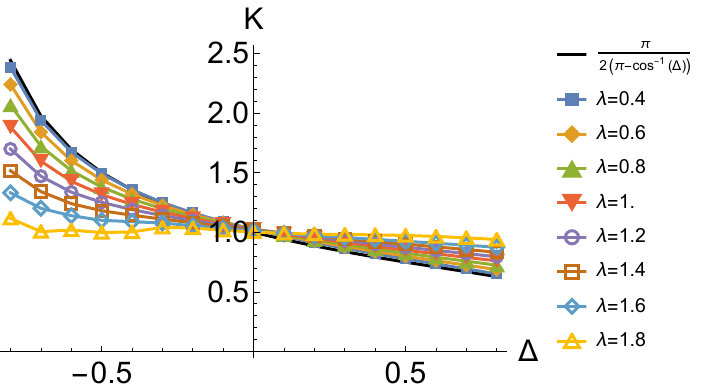}
    \includegraphics[width=0.43\textwidth]{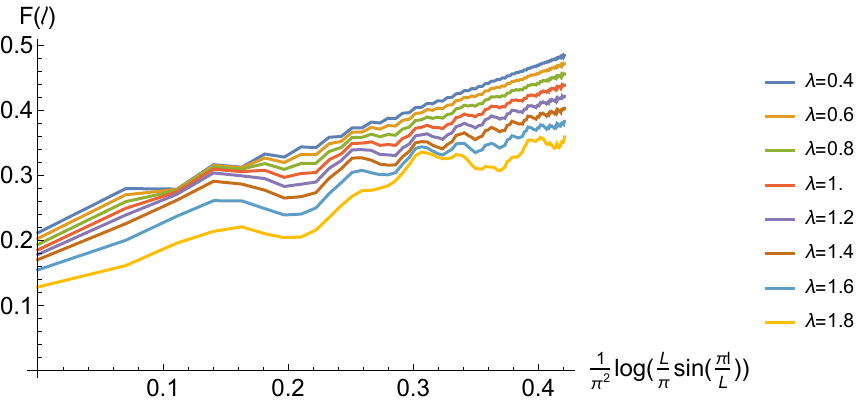}
    \caption{Upper panel: Luttinger parameter $K$ extracted from subsystem charge fluctuations, as a function of the interaction strength $\Delta$, for various values of the quasiperiodic potential strength $\lambda$. Note that as $\lambda$ is increased the curves get increasingly flat, at the noninteracting value $K = 1$. Lower panel: dependence of charge fluctuations on subsystem size for interactions $\Delta = - 0.5$ and various values of $\lambda$. DMRG simulations were done with $L=200$ and a maximum bond dimension of 200. Data is averaged over 20 phase realizations.}
    \label{K_chargefluct}
\end{figure}

The results for the charge fluctuations are shown in Fig.~\ref{K_chargefluct}. The charge fluctuations are logarithmic in $\ell$, as expected for a Luttinger liquid. Extracting $K$ from the coefficient of the logarithm, we find that $K \to 1$---the noninteracting value---as $\lambda$ is increased at fixed interaction strength $\Delta$. As one might expect, the onset of the clear logarithmic scaling regime is pushed out to larger $\ell$ as we increase $\lambda$. 
In~\cite{suppmat} we also consider a separate diagnostic, namely the two-point correlator of the ``spins'' related to the fermions of Eq.~\eqref{eq1} by a Jordan-Wigner transformation. This diagnostic is \emph{a priori} unrelated to the charge fluctuations, but in a Luttinger liquid, these two observables are tightly related. Consistent estimates of $K$ from these two approaches are a consistency check that the system is a Luttinger liquid. 
%

These numerical observations lead to the conclusions that (1)~the system is a Luttinger liquid throughout the delocalized phase of the Aubry-Andr\'e model, and (2)~the Luttinger parameter is renormalized toward its noninteracting value as the quasiperiodic potential is increased. These observations are consistent with the qualitative picture presented in the preceding section. 


\emph{Kane-Fisher effect: self-consistent approach}.---The previous section presented numerically exact calculations for the case of nearest-neighbor interactions. We now turn to an approximate but versatile approach that sheds light on the mechanism for Luttinger-parameter suppression, and also lets us extend our analysis to models with tunable-range interactions. In this section we introduce the basic idea behind the method for clean systems; the next section applies the method to quasiperiodic systems.

The basic tool we will use to diagnose correlations is the Kane-Fisher effect~\cite{PhysRevLett.68.1220}---i.e., the renormalization of tunneling across a localized impurity due to interactions. At zero temperature and at incident energies $E$ close to the Fermi energy $E_F$, the renormalized transmission coefficient across the impurity scales as $T(E) \sim |E-E_F|^{2(K^{-1}-1)}$. Thus, tunneling is parametrically suppressed (enhanced) for repulsive (attractive) interactions. 
Although the Kane-Fisher effect is normally studied using bosonization techniques that do not generalize to our setting, there is an alternative weak-coupling formulation due to Yue et al.~\cite{PhysRevB.49.1966} that is more versatile. In this treatment, one begins with the observation that a static impurity in a Fermi gas creates Friedel oscillations in the density profile of the surrounding gas. These Friedel oscillations renormalize the effective scattering shift experienced by electrons farther away from the impurity---thus further renormalizing the Friedel oscillations, and so forth. Solving self-consistently for the tunneling across the impurity one recovers the Kane-Fisher effect. 
%

\begin{figure*}[tb]
     \centering
         \centering
         \includegraphics[width=0.32\textwidth]{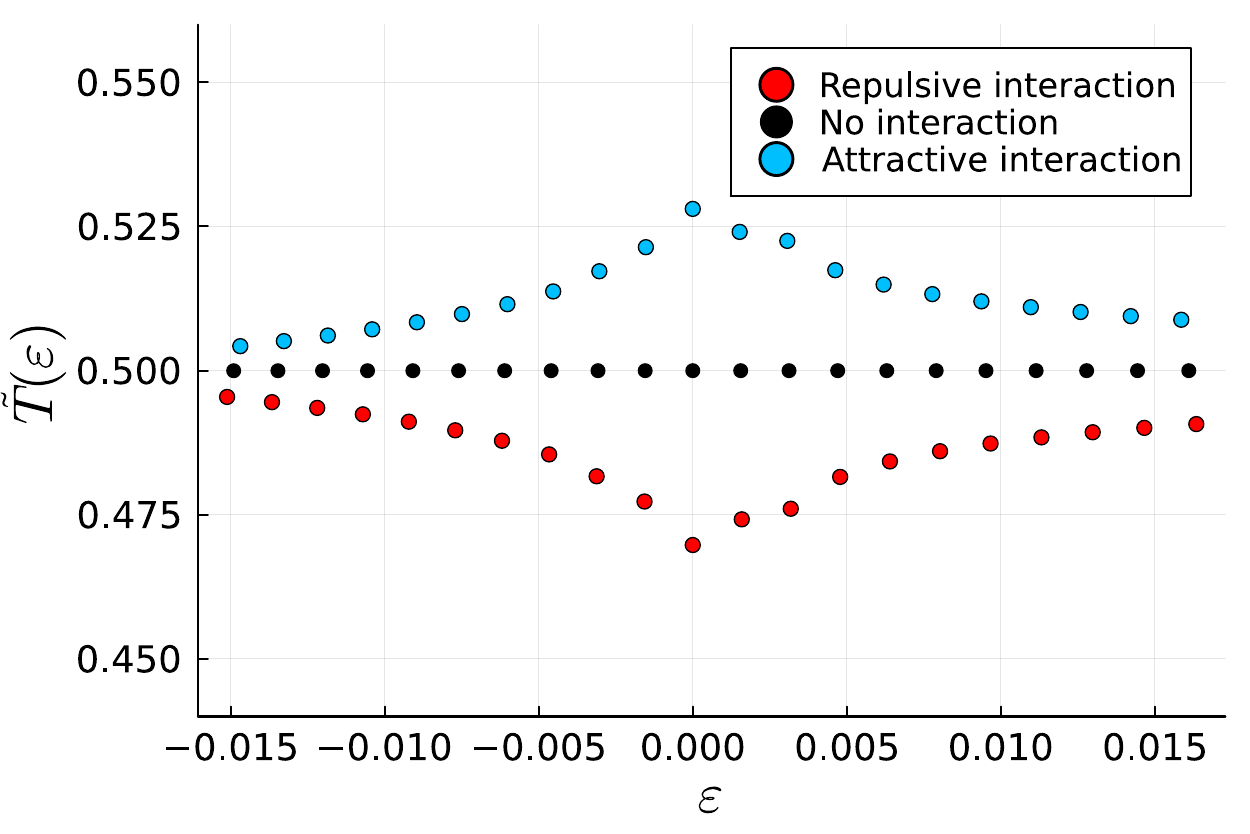}
%
         \includegraphics[width=0.32\textwidth]{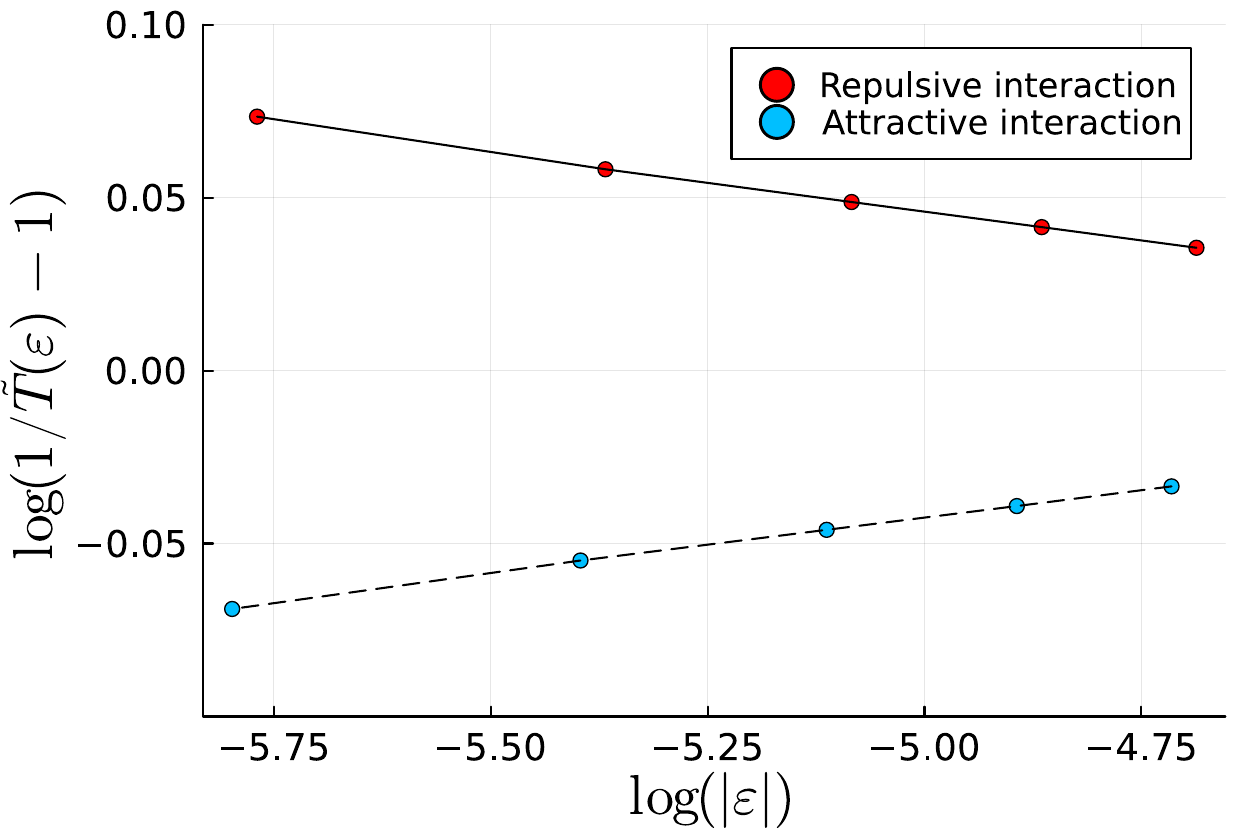}
        \includegraphics[width = 0.32\textwidth]{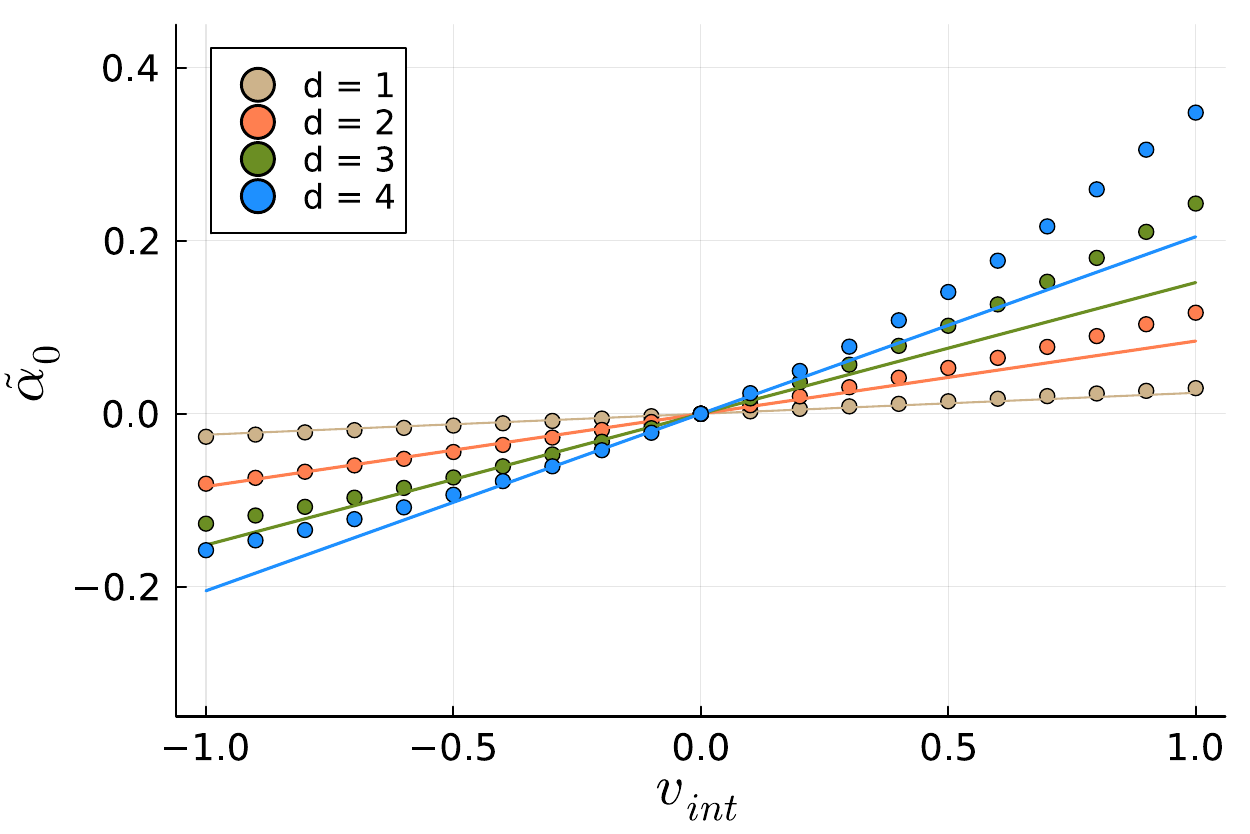}
        \caption{Data on transmission coefficients in the clean system, computed for system size $N = 2501$ and filling fraction $f = 0.1$. Left: Transmission coefficient as a function of energy measured from the Fermi level, for $v_{\mathrm{int}} = \pm 0.6$ and without interactions. Middle: same data but on a log-log plot, showing the nonanalytic behavior of the transmission coefficient near $E_F$. Right: Exponent $\tilde\alpha$ (dots) compared with the continuum analytic prediction $\alpha$ [Eq.~\eqref{equation:analytic_alpha}], as a function of interaction strength, for various choices of the interaction range. Discrepancies at larger interaction strength are expected as lattice dispersion effects become significant .}
        \label{fig:clean_transmission}
\end{figure*}

To explore the effect of quasiperiodic modulations on the Kane-Fisher effect, we first self-consistently solve for the ground state in the presence of the impurity, in the Hartree-Fock approximation, and then numerically solve for tunneling across the impurity. It might seem surprising that a mean-field approximation like Hartree-Fock can capture a correlation-dominated quantity like the Luttinger parameter. Intuitively, the reason why this is possible is that the way the electrons avoid the impurity in the mean-field treatment is essentially analogous to the way they avoid each other due to correlations. In any case, we note that the predicted power-law divergences are clearly seen in the self-consistent Hartree-Fock numerical solution (Fig.~\ref{fig:clean_transmission}).

Before turning to our numerical results, we briefly summarize the main results of Ref.~\cite{PhysRevB.49.1966}.
This work linearizes the dispersion around the Fermi surface, projecting to states with energies within $D_0 = v_F / d$ of $E_F$. 
Solving self-consistently for the potential~\cite{suppmat} gives the result for the transmission coefficient \cite{PhysRevB.49.1966}
\begin{equation}
    {T}(\epsilon) = \frac{{T}_0 |\epsilon / D_0|^{2\alpha}}{{R}_0 + {T}_0 |\epsilon / D_0|^{2\alpha}},
    \label{equation:analytic_t}
\end{equation}
where $\epsilon = E - E_F$ and ${T}_0 = |t_0|^2$ and ${R}_0 = 1 - {T}_0$ are the bare transmission and reflection coefficients. The dimensionless parameter $\alpha$ is defined by \cite{PhysRevB.49.1966}
\begin{eqnarray}
&&    \alpha = \alpha_{\mathrm{exchange}} - \alpha_{\mathrm{Hartree}} \nonumber \\
&&  \alpha_{\mathrm{exchange}} = \frac{\tilde{V}(0)}{2\pi v_F}, \quad \alpha_{\mathrm{Hartree}} = \frac{\tilde{V}(2k_F)}{2\pi v_F}.
    \label{equation:analytic_alpha}
\end{eqnarray}
Here $\alpha = K^{-1} - 1$. 
In the continuum limit, the exchange and Hartree terms have opposite effects on the transmission coefficient. In the case of a repulsive (attractive) interaction, the exchange term decreases (increases) the transmission coefficient compared to its bare value, while the Hartree term increases (decreases) it. For zero-range interactions, the exchange and Hartree terms cancel exactly, as they must because contact interactions have no effect on spinless fermions.

We now describe our numerical method for calculating transmission coefficients on the lattice, focusing first on the case without a quasiperiodic potential. We will work with Eq.~\eqref{eq1} with a finite-range Gaussian interaction Eq.~\eqref{intrange}. 
We perform a mean-field Hartree-Fock decoupling of the interaction term: $V_\mathrm{int}(i,j) n_i n_j \approx V_{H}(i)c_{i}^{\dagger}c_{i} + \sum_{i \neq j }V_{\rm ex}(i, j)c_{i}^{\dagger}c_{j}$, 
where $V_{H}(i) = \sum_{j \neq i}V_{\rm int}(|i - j|)\braket{c_{j}^{\dagger}c_{j}}$ and $V_{\rm ex}(i, j) = -V_{\rm int}(|i - j|)\braket{c_{j}^{\dagger}c_{i}}$.

To compute the transmission coefficient we proceed as follows. First, we solve the Hartree-Fock problem fully self-consistently for the ground state. Next, we treat the dynamics of wavepacket transmission under this self-consistent quadratic Hamiltonian. In the clean case, the transmission coefficient can be computed using scattering theory. Once the quasiperiodic potential is turned on, we do not have a simple expression for the asymptotic states. Therefore, to calculate the transmission coefficient, we construct a wavepacket localized in real space to one side of the barrier, and sharply localized in energy space at some energy $E$ displaced by $\epsilon$ from $E_F$. We allow the wavepacket to evolve in time and scatter off the barrier, defining the transmission coefficient $\tilde{T}(\epsilon)$ as the modulus squared of the portion of the wavepacket that passes through the barrier to the other side. 

Our results for the clean case are summarized in Fig.~\ref{fig:clean_transmission}. To minimize lattice effects we work at a filling $f = 0.1$. The left panel shows the transmission coefficient as a function of energy (measured with reference to $E_F$), showing a dip (peak) for repulsive (attractive) interactions. Note that in the noninteracting case the transmission coefficient is smooth across $E_F$. The middle panel of Fig.~\ref{fig:clean_transmission} shows that the transmission diverges/vanishes as a power law, consistent with Eq.~\eqref{equation:analytic_t}. 
%
%
From this power-law dependence one can extract a numerical exponent $\tilde\alpha$. 
%
This is shown in the right panel of Fig.~\ref{fig:clean_transmission}. 
In the thermodynamic limit, for weak enough interactions and low enough filling, we expect that the numerical value $\tilde\alpha$ should approach the continuum analytic value of $\alpha$~\eqref{equation:analytic_alpha}.
Our data are consistent with this. For stronger interactions, it is no longer quantitatively accurate to linearize the lattice dispersion around the Fermi surface, so $\tilde{\alpha}$ deviates from $\alpha$. (Of course, the Hartree-Fock framework also ceases to be quantitatively accurate.) 





\begin{figure*}[bt]
    \centering
    \includegraphics[width=0.32\textwidth]{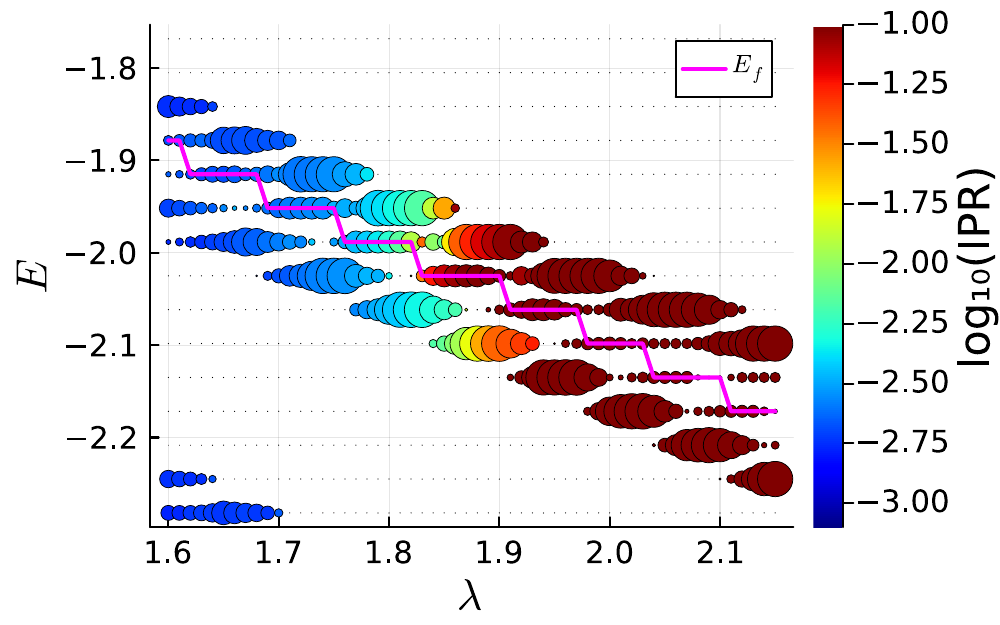}
    \includegraphics[width=0.32\textwidth]{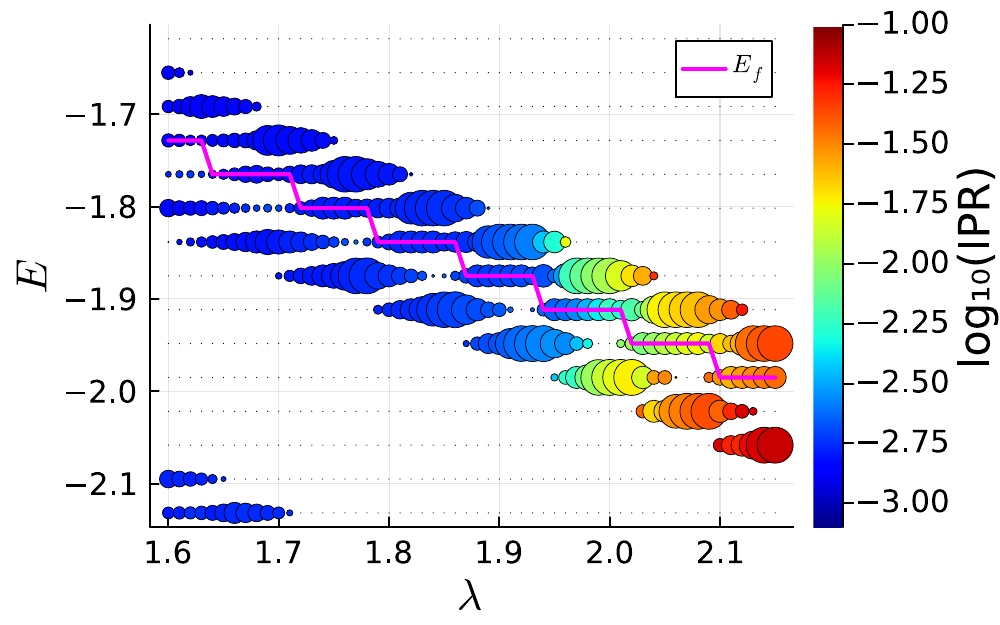}
    \includegraphics[width=0.32\textwidth]{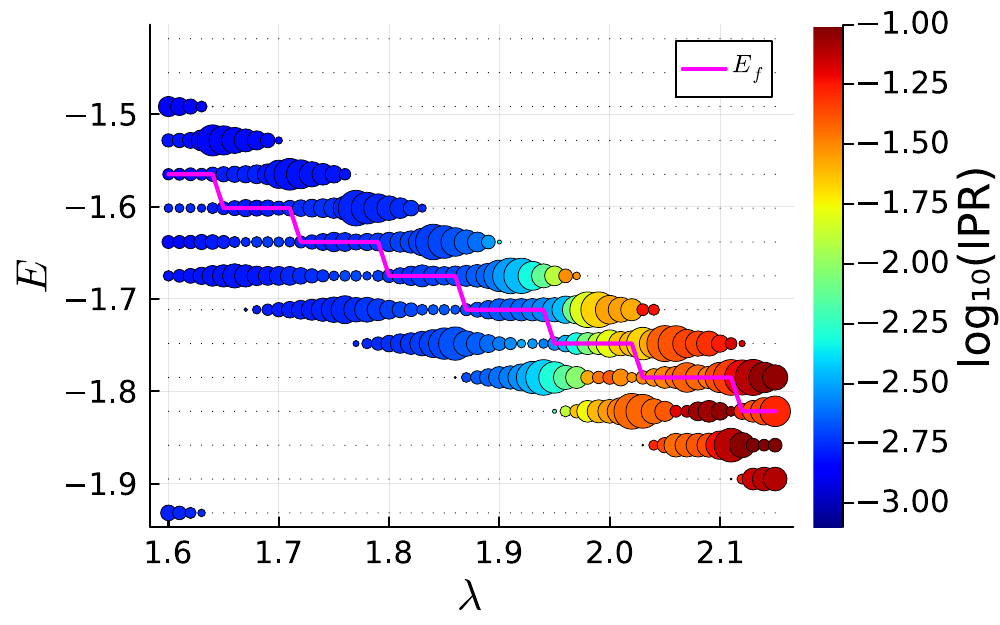}
    \caption{Localization transition as a function of interactions, for energies near the Fermi level (marked as a red line), for attractive, noninteracting, and repulsive cases, with interaction range $d = 4$. The inverse participation ratio (IPR) of a normalized single-particle state with real-space amplitudes $\psi_i$ on a lattice is defined as $\sum\nolimits_i |\psi_i|^4$. Strictly speaking, localization is diagnosed by how the IPR scales with system size, but at this large size $N = 2001$ a large IPR is a good proxy for localization. The localization transition clearly drifts to smaller $\lambda$ in the attractive case; the trend in the repulsive case is ambiguous. Data is averaged over 100 phase realizations.}
    \label{fig:localization}
\end{figure*}

\emph{Kane-Fisher effect in quasiperiodic systems}.---
Having checked that our self-consistent method reproduces the Kane-Fisher effect in clean systems, we turn to quasiperiodic systems. Quasiperiodic systems have nontrivial spatial dependence of the self-consistent Hartree and exchange potentials even in the absence of an impurity. These effects renormalize the band structure and also shift the localization transition. These renormalization effects were previously studied in Refs.~\cite{mastropietro2015localization, PhysRevLett.126.040603, PhysRevResearch.3.033257, PhysRevLett.129.103401}, and generally give rise to a mobility edge~\cite{PhysRevLett.129.103401, PhysRevA.80.021603}. Intuitively,
attractive interactions enhance the localization of low-energy states via self-trapping, while repulsive interactions suppress localization by screening the quasiperiodic potential. These observations are broadly consistent with our numerical study of the inverse participation ratio (IPR) as a function of disorder and energy (Fig.~\ref{fig:localization}).

We now explore the Kane-Fisher effect in the regime where states near the Fermi energy are clearly delocalized. 
Following previous studies of impurity models in quasiperiodic potentials~\cite{PhysRevB.100.165116, PhysRevB.106.165123}, we work at a filling factor $f = 0.31$ at which the Fermi level lies well inside a miniband throughout the delocalized phase. 
We proceed exactly as above to compute the transmission coefficient, and extract a power law from its singular behavior near $E_F$.
One subtlety that arises in the quasiperiodic case is that the tunneling depends strongly on the phase of the quasiperiodic potential. In order to isolate the effect of interactions on the transmission coefficient at any given combination of $\lambda$ and $\theta$ and to facilitate the observation of the ways in which attractive and repulsive interactions renormalize the bare transmission coefficient, we tune the barrier strength to fix the bare transmission coefficient for the wavepacket at the Fermi surface to the arbitrary value of $0.5 \pm 0.01$.

\begin{figure}[tb]
    \centering
    \includegraphics[width=0.45\textwidth]
    {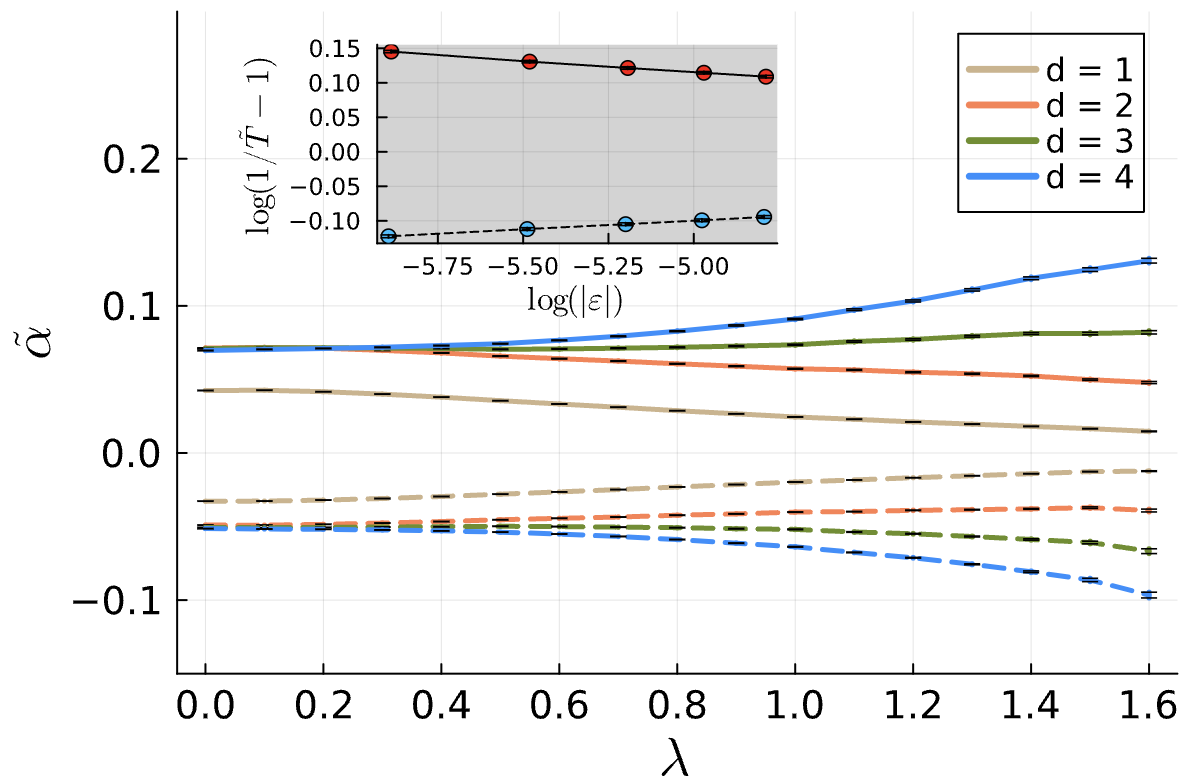}
    \caption{Exponents $\tilde{\alpha}$ characterizing the Kane-Fisher effect in systems with increasing quasiperiodic potential strength $\lambda$. Data is computed for system size $N = 2501$, filling fraction $f = 0.31$, $v_{int} = \pm 0.6$, and averaged over 100 phase realizations. Solid lines: $\tilde{\alpha}$ from repulsive interactions; dashed lines: $\tilde{\alpha}$ from attractive interactions. Inset: power-law relationship between $\epsilon$ and $1/\tilde{T}(\epsilon) - 1$ for system with $\lambda = 1.5$ and interaction range $d = 1$. }
    \label{fig:incr_vqp}
\end{figure}

Our results are shown in Fig.~\ref{fig:incr_vqp}. We find that for a short-ranged interaction with $d = 1$, increasing $\lambda$ pushes $\tilde\alpha$ toward zero. Long-ranged interactions have the opposite effect, with $\alpha$ flowing away from zero as we increase $\lambda$. Past $\lambda \approx 1.7$, the transmission coefficient no longer behaves as a power-law in $|E - E_F|$ in the energy range we are able to resolve for the available system sizes of size $\alt 3000$ sites. This behavior is illustrated in~\cite{suppmat1}. 

\begin{figure}[bt]
    \centering
        \includegraphics[width=0.45\textwidth]
        {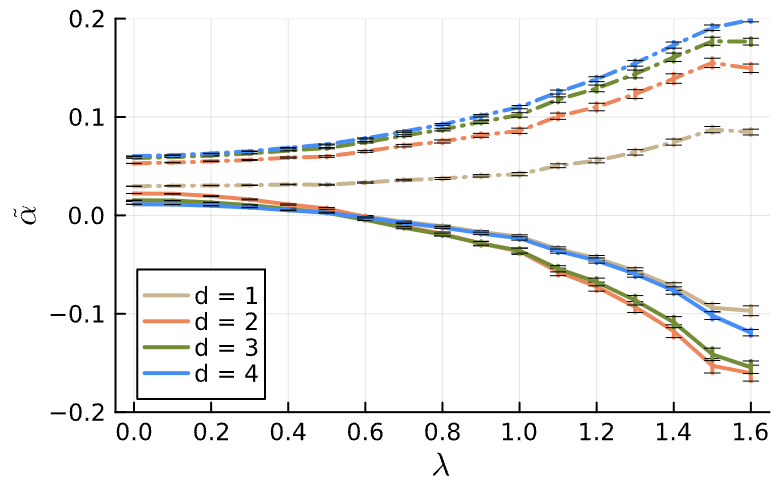}

    \caption{Hartree (exchange) exponents $\tilde{\alpha}$ characterizing Hartree (exchange) contribution to the Kane-Fisher effect in systems with increasing quasiperiodic potential strength $\lambda$. Data is computed for system size $N = 2501$, filling fraction $f = 0.31$, repulsive interaction $v_{int} = 0.6$, and averaged over 100 phase realizations. Solid lines: Hartree exponent. Dash-dot lines: Exchange exponent.}
\label{fig:incr_vqp_hartree}
\end{figure}

To understand the origin of these effects, it is instructive to separate out the effects due to the Hartree and exchange interaction terms. To do this, we first solve for the self-consistent potential including both terms. Then to compute the Hartree (exchange) contribution to the transmission coefficient, we evaluate wavepacket dynamics in a quadratic Hamiltonian where the exchange (Hartree) term is artificially set to zero. From this we again extract an exponent, which we call the Hartree (exchange) exponent (Fig.~\ref{fig:incr_vqp_hartree}). The exchange exponent behaves as one would expect, growing stronger with increasing $\lambda$ as the bands grow flatter. The behavior of the Hartree term is more subtle. At filling fraction $0.31$, the Bloch states are rather far from continuum plane waves, so the Hartree term is not well described by simply taking the Fourier transform of the interaction potential. In the clean limit, at this filling, we find that the Hartree and exchange terms in fact have the same sign. Upon increasing $\lambda$, the sign of the Hartree effect flips. The strength of this term is also non-monotonic with interaction range: for long-range interactions the term is increasingly suppressed, as one would expect (a smoothly varying potential has no effect on scattering). However, for short range potentials, the Hartree term becomes increasingly strong and negative as $\lambda$ is increased. This is the origin of the suppression of interaction effects: as remarked in the introduction, the interactions projected onto the miniband at the Fermi level are increasingly contact-like as the minibands become sparser, so the Hartree and Fock terms cancel as they would for contact interactions.



\emph{Discussion}.---
This work explored how Luttinger-liquid behavior emerges in the low-energy limit of interacting quasiperiodic systems. We found clear evidence by two different methods that Luttinger-liquid behavior does emerge, albeit with a strongly renormalized Luttinger parameter. The Luttinger parameter exhibits a counterintuitive dependence on the strength of the quasiperiodic potential $\lambda$: for short-range interactions, as $\lambda$ is tuned toward the localization transition, the Luttinger parameter approaches that of the free system, although the bands get very flat. We identified the origin of this phenomenon as a cancellation between Hartree and Fock terms that becomes more pronounced at larger $\lambda$. 
This cancellation occurs because the characteristic spatial scale of the miniband at the Fermi level grows (with correlation length $\xi \sim 1/|\lambda_c - \lambda|$) with increasing $\lambda$. 

Our conclusions are consistent with (and complementary to) those found by Ref.~\cite{goncalves} exploring interaction effects at the Aubry-Andr\'e critical point: while that work focuses on the critical point itself, we show how this regime emerges coming from the Luttinger liquid. 
Ref.~\cite{goncalves} concluded, based on a scaling argument, that finite-range interactions are always irrelevant at the Aubry-Andr\'e critical point. 
This would imply that the curves in Fig.~\ref{fig:incr_vqp} are in fact non-monotonic: very close to the transition, we expect the effective Luttinger parameter to dip again toward $K = 1$. 
We do not see this numerically, and there are at least two plausible explanations for why. The first is that the Luttinger parameter eventually dips back down, but in a parameter regime where we are unable to see a clean Kane-Fisher effect. 
A second possibility is that at finite (as opposed to infinitesimal) interaction strengths, interactions renormalize the localization problem sufficiently to alter the nature of the Aubry-Andr\'e transition (e.g., the fractal structure of energy levels). 
This second possibility is less exotic than it sounds. If one considers a two-dimensional phase diagram with interaction strength and quasiperiodicity as the axes, the critical behavior found in Ref.~\cite{goncalves} must change to the behavior found using Luttinger-liquid theory~\cite{roux2008quasiperiodic} somewhere along the phase boundary, and it is natural for this location to depend on interaction range. 
Exploring this question in more detail, and developing a complete theory of this unusual critical phenomenon, is an interesting challenge for future work.

%

The framework we have used also naturally extends to two-dimensional systems. Quasiperiodic systems in higher dimensions exhibit a richer phase diagram at the noninteracting level, with ballistic, diffusive, and localized phases~\cite{devakul2017anderson, PhysRevB.101.014205}. The analog of the Kane-Fisher effect in two-dimensions is the Altshuler-Aronov suppression of the density of states~\cite{PhysRevLett.44.1288}, which can again be understood in terms of renormalized Friedel oscillations. It would be interesting to explore these effects in the quasiperiodic case.

\begin{acknowledgments}
We thank David Huse and Jed Pixley for useful conversations. This work was supported by NSF Grants No.  DMR-2103938 (T.J.V., S.G.) and DMR-2104141 (R.V.).
\end{acknowledgments}

\appendix

\begin{widetext}

\section*{Supplemental Material}

In this document we provide additional numerical results and details on the following points: (1)~Additional DMRG data supporting Luttinger-liquid behavior, (2)~A brief review of the derivation in Ref.~\cite{PhysRevB.49.1966} of the continuum Kane-Fisher effect, and (3)~data on how the Kane-Fisher effect breaks down as one approaches the transition to localization. 

\section{Luttinger-parameter data from correlation functions}

In the main text, we showed data extracting the Luttinger parameter from the dependence of charge variance on subsystem size. Here we present data extracting the Luttinger parameter from a different observable, namely the correlation function of the ``spin'' operator $G_{+-}(x) \equiv \langle S^+(x) S^-(0)\rangle$, where $S^+(x) = \exp\left(-i\pi \sum_{y < x} c^\dagger(y) c(y) \right) c^\dagger(x)$. In a Luttinger liquid we expect $G_{+-}(x) \sim |x|^{-1/(2K)}$ up to oscillating terms. This gives us an independent method for extracting the Luttinger parameter and confirming that the system is a Luttinger liquid (i.e., the correlations do decay as a power law). The relevant data are shown in Fig.~\ref{luttcorr}. The data extracted this way are in close agreement with those presented in the main text, which were extracted from the charge variance.

\begin{figure}
    \centering
    \includegraphics[width=0.32\linewidth]{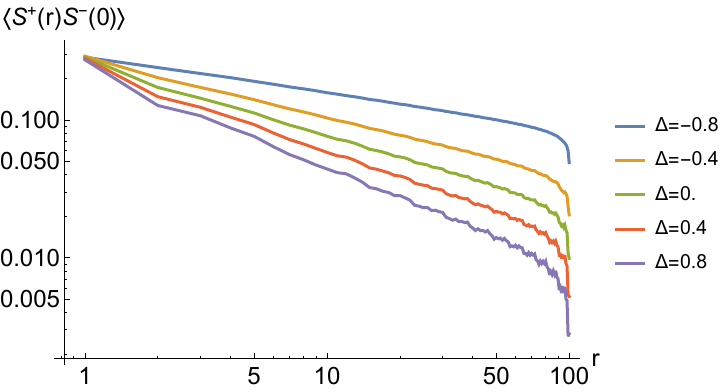}
    \includegraphics[width=0.32\linewidth]{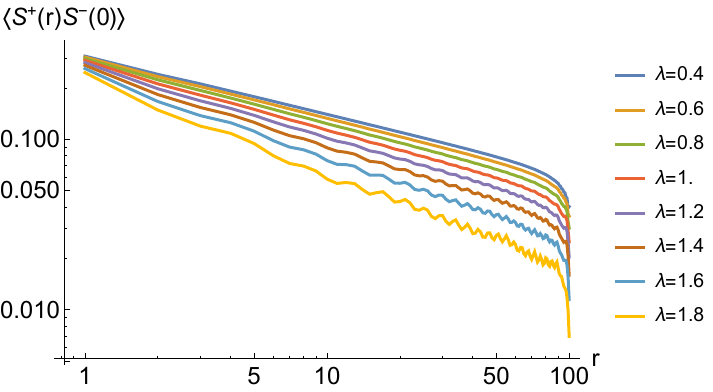}    \includegraphics[width=0.32\linewidth]{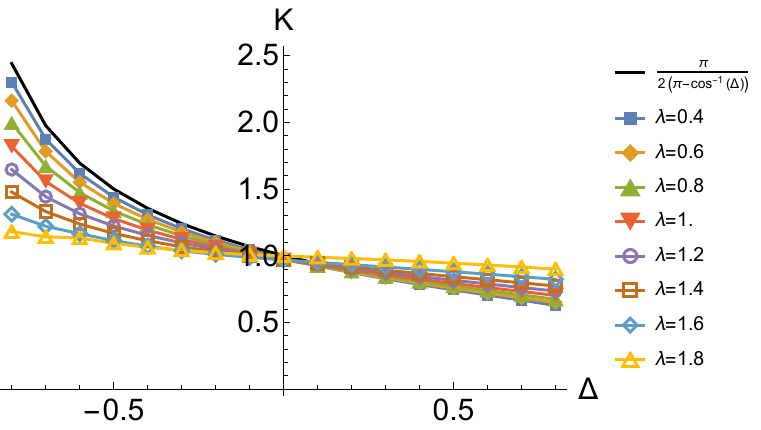}
    \caption{Data on the correlation function $G_{+-}(r)$ with distance $r$ extracted using DMRG on a system of size $100$. Left: variation with interaction strength $\Delta$ at fixed $\lambda = 1$. Center: variation with $\lambda$ at fixed $\Delta = - 0.5$. Right: Luttinger parameter $K$ extracted from these power laws. These results for the Luttinger parameter are consistent with those presented in the main text, where they were extracted from the charge variance. DMRG simulations were done with $L=200$ and a maximum bond dimension of 200. Data is averaged over 20 phase realizations.}
    \label{luttcorr}
\end{figure}

\section{Review of the continuum argument in Ref.~\cite{PhysRevB.49.1966}}

In this section, we review the scattering of spinless interacting electrons from a barrier $U(x)$ localized at the origin of a continuous, infinite one-dimensional system. In the absence of interactions, the electronic eigenfunctions are 
\begin{equation}
   \phi_{k}(x) = \frac{1}{\sqrt{2\pi}}
   \[ \begin{cases} 
      e^{ikx} + r_0 e^{-ikx} & x < -a \\
      t_0 e^{ikx} & a < x \\
   \end{cases}
\],
\end{equation}
\begin{equation}
   \phi_{-k}(x) = \frac{1}{\sqrt{2\pi}}
   \[ \begin{cases} 
      t_{0}' e^{-ikx} & x < -a \\
      e^{-ikx} + r_{0}'e^{ikx} & a < x \\
   \end{cases}
\].
\end{equation}
$a$ denotes the spatial extent of the barrier. In the presence of an interaction $V(|x - y|)$, one can use Hartree-Fock theory to describe the way in which electrons are affected by the other electrons in the Fermi sea. In this theory, each electron is subject to a Hartree potential $V_H(x)$ and an exchange potential $V_{ex}(x, y)$, defined by
\begin{equation}
    V_H(x) = \int dy V(|x - y|) n(y),
\end{equation}
\begin{equation}
    V_{ex}(x, y) = -V(|x - y|) \sum_{|q| < k_F}\phi_{q}^{*}(y)\phi_{q}(x),
\end{equation}
where $n(y) = \sum_{|q| < k_F}\phi_{q}^{*}(y)\phi_{q}(y)$.

The Hartree-Fock equations determining the single-electron wavefunctions are 
\begin{equation}
\begin{split}
    [-\frac{1}{2m}\frac{d^2}{dx^2} &+ U(x) + V_H(x)]\phi_{k}(x) \\ &+ \int dy V_{ex}(x, y)\phi_{k}(y) = E\phi_{k}(x)
\end{split}
\label{equation:hartreefock}
\end{equation}

Naively, one can calculate modified transmission amplitudes $t_k$ to leading order by obtaining the Hartree and exchange potentials in the first Born approximation, substituting them into Eq. \ref{equation:hartreefock}, using the method of Green's functions to obtain the asymptotic form of the single-electron wavefunctions, and reading off the coefficient $t_k$:  
\begin{equation}
    \phi_{k}(x) = \frac{1}{\sqrt{2\pi}} t_{k}e^{ikx}, \quad \quad x \rightarrow \infty.
\end{equation}
However, corrections to $t_0$ due to the Hartree and exchange potentials at first order in perturbation theory contain logarithmic divergences as $k \rightarrow k_F$ \cite{PhysRevB.49.1966}. In fact, these divergences are present at any finite order in perturbation theory. One way to proceed is to sum the most divergent terms at all orders in perturbation theory, which can be accomplished via renormalization in energy space. Performing this renormalization via a poor man's scaling approach yields the results quoted in the main text. 

\section{Breakdown of Kane-Fisher effect at large $\lambda$}

In the main text we presented data on the numerical exponent $\tilde\alpha$ characterizing the Kane-Fisher effect, in the regime of quasiperiodic potential $\lambda$ for which this was well-defined. In Fig.~\ref{breakdown} we provide an example of how this structure breaks down as we increase $\lambda$ keeping the other parameters fixed. 

\begin{figure}
    \centering
    \includegraphics[width=0.45\linewidth]{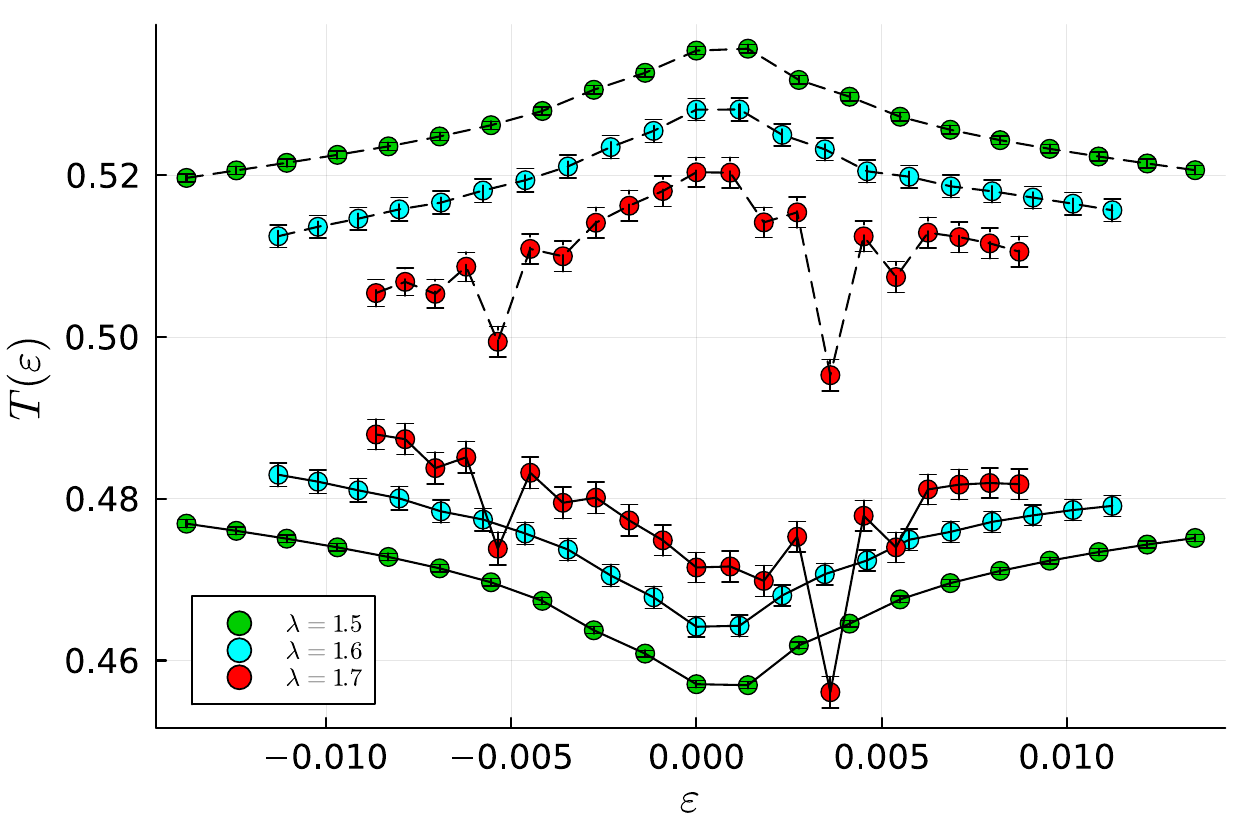}
    \includegraphics[width=0.45\linewidth]{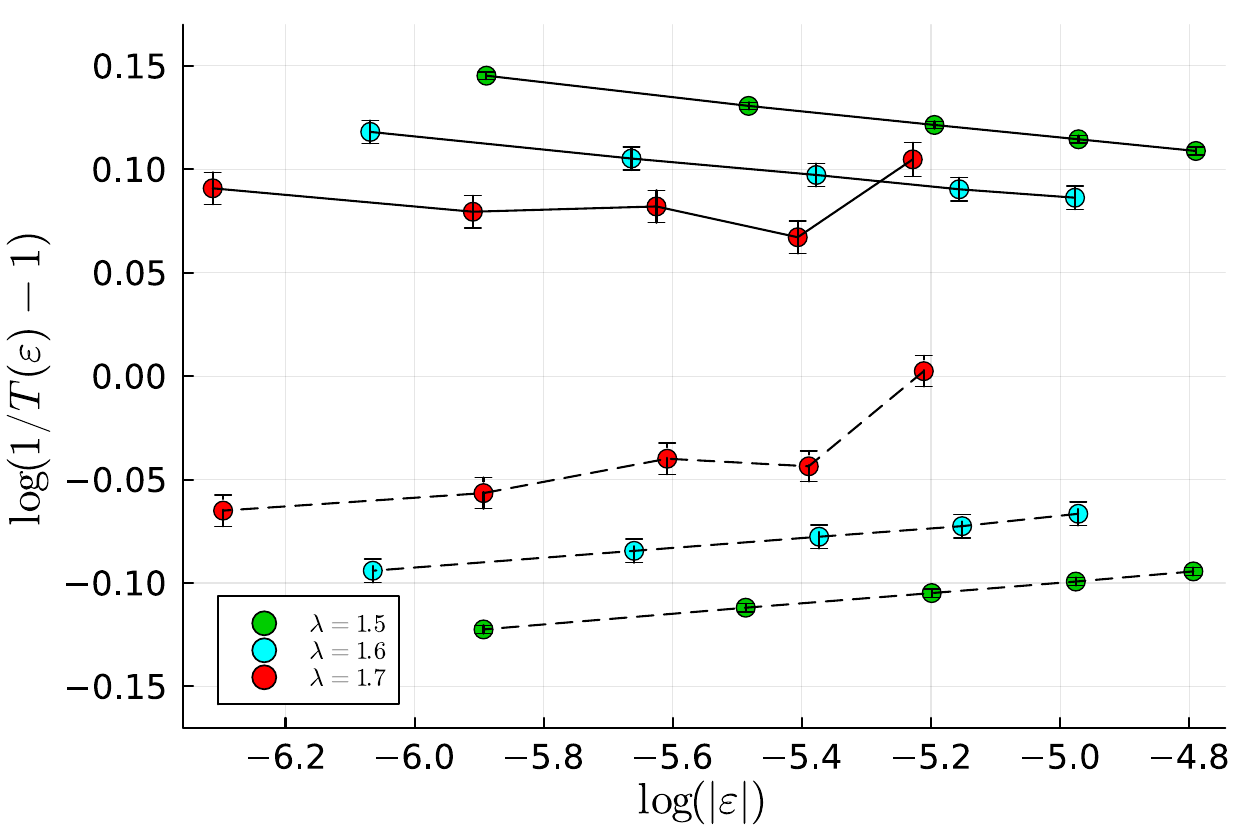}
    \caption{Behavior of transmission amplitudes near the localization transition. Left: behavior of the transmission amplitude near the Fermi energy for various $\lambda$ at $v_{\mathrm{int}} = 0.6$. Right: same data but on a log-log plot, showing the breakdown of the Kane-Fisher power law (except presumably at inaccessibly small energies).}
    \label{breakdown}
\end{figure}

\medskip

\end{widetext}

\bibliography{mybibliography}

\end{document}